\documentclass[final,3p,times]{elsarticle}

\usepackage{graphicx}
\usepackage{amssymb}
\usepackage{amsmath}
\usepackage{hyperref}
\usepackage[usenames]{color}

\journal{Computer Physics Communications}

\begin{document}

\begin{frontmatter}

\title{OpenMP GNU and Intel Fortran programs for solving the time-dependent Gross-Pitaevskii equation}

\author[col]{Luis E. Young-S.}
\ead{luis.young@usantoto.edu.co}

\author[bdu]{Paulsamy Muruganandam}
\ead{anand@cnld.bdu.ac.in}

\author[ift]{Sadhan K. Adhikari}
\ead{adhikari@ift.unesp.br}

\author[scl]{Vladimir Lon\v{c}ar}
\ead{vladimir.loncar@ipb.ac.rs}

\author[scl]{Du\v{s}an Vudragovi\'{c}}
\ead{dusan.vudragovic@ipb.ac.rs}

\author[scl]{Antun Bala\v{z}\corref{author}}
\ead{antun.balaz@ipb.ac.rs}

\cortext[author]{Corresponding author.}
\address[col]{Departamento de Ciencias B\'asicas, Universidad Santo Tom\'as, 150001 Tunja, Boyac\'a, Colombia}
\address[bdu]{Department of Physics, Bharathidasan University, Palkalaiperur Campus, Tiruchirappalli -- 620024, Tamil Nadu, India}
\address[ift]{Instituto de F\'{\i}sica Te\'{o}rica, UNESP -- Universidade Estadual Paulista, 01.140-70 S\~{a}o Paulo, S\~{a}o Paulo, Brazil}
\address[scl]{Scientific Computing Laboratory, Center for the Study of Complex Systems, Institute of Physics Belgrade, University of Belgrade, Serbia}

\begin{abstract}
We present Open Multi-Processing (OpenMP) version of Fortran 90 programs for solving the Gross-Pitaevskii (GP) equation for a Bose-Einstein condensate in one, two, and three spatial dimensions, optimized for use with GNU and Intel compilers. We use the split-step Crank-Nicolson algorithm for imaginary- and real-time propagation, which enables efficient calculation of stationary and non-stationary solutions, respectively. The present OpenMP programs are designed for computers with multi-core processors and optimized for compiling with both commercially-licensed Intel  Fortran and popular free open-source GNU Fortran compiler. The programs are easy to use and are elaborated with helpful comments for the users. All input parameters are listed at the beginning of each program. Different output files provide physical quantities such as energy, chemical potential, root-mean-square sizes, densities, etc. We also present speedup test results for new versions of the programs.
\end{abstract}

\begin{keyword}
Bose-Einstein condensate; Gross-Pitaevskii equation; Split-step Crank-Nicolson scheme;
Intel and GNU Fortran programs; Open Multi-Processing; OpenMP; Partial differential equation

\PACS 02.60.Lj; 02.60.Jh; 02.60.Cb; 03.75.-b
\end{keyword}

\end{frontmatter}

\begin{small}
\noindent
{\bf New version program summary}

\noindent\\
{\em Program title:} BEC-GP-OMP-FOR software package, consisting of: (i) imag1d-th, (ii) imag2d-th, (iii) imag3d-th, (iv) imagaxi-th, (v) imagcir-th, (vi) imagsph-th, (vii) real1d-th, (viii) real2d-th, (ix) real3d-th, (x) realaxi-th, (xi) realcir-th, (xii) realsph-th.

\noindent\\
{\em Program files doi:} \href{http://dx.doi.org/10.17632/y8zk3jgn84.2}{http://dx.doi.org/10.17632/y8zk3jgn84.2}\\
{\em Licensing provisions:} Apache License 2.0\\
{\em Programming language:} OpenMP GNU and Intel Fortran 90.\\
{\em Computer:} Any multi-core personal computer or workstation with the appropriate OpenMP-capable Fortran compiler installed. \\
{\em Number of processors used:} All available CPU cores on the executing computer. \\
{\em Journal reference of previous version:} Comput. Phys. Commun. \textbf{180} (2009) 1888; \textit{ibid.} \textbf{204} (2016) 209.\\
{\em Does the new version supersede the previous version?:} Not completely. It does supersede previous Fortran programs from both references above, but not OpenMP C programs
from Comput. Phys. Commun. \textbf{204} (2016) 209.

\noindent\\
{\em Nature of problem:}
The present Open Multi-Processing (OpenMP) Fortran programs, optimized for use with commercially-licensed Intel Fortran and free open-source GNU Fortran compilers, solve the time-dependent nonlinear partial differential GP equation for a trapped Bose-Einstein condensate in one (1d), two (2d), and three (3d) spatial dimensions for six different trap symmetries: axially and radially symmetric traps in 3d, circularly symmetric traps in 2d, fully isotropic (spherically symmetric) and fully anisotropic traps in 2d and 3d, as well as 1d traps, where no spatial symmetry is considered.

\noindent\\
{\em Solution method:}
We employ the split-step Crank-Nicolson algorithm to discretize the time-dependent GP equation in space and time. The discretized equation is then solved by imaginary- or real-time propagation, employing adequately small space and time steps, to yield the solution of stationary and non-stationary problems, respectively.

\noindent\\
{\em Reasons for the new version:}
Previously published Fortran programs \cite{bec2009,bec2016} have now become popular tools \cite{uca} for solving the GP equation. These programs have been translated to the C programming language \cite{bec2012} and later extended to the more complex scenario of dipolar atoms \cite{dbec2015}. Now virtually all computers have multi-core processors and 
some have motherboards with more than one physical computer processing unit (CPU), which may increase the number of available CPU cores on a single computer to several tens. The C programs have been adopted to be very fast on such multi-core modern computers using general-purpose graphic processing units (GPGPU) with Nvidia CUDA and computer clusters using Message Passing Interface (MPI) \cite{dbec2016}. Nevertheless, previously developed Fortran programs are also commonly used for scientific computation and most of them use a single CPU core at a time in modern multi-core laptops, desktops, and workstations. Unless the Fortran programs are made aware and capable of making efficient use of the available CPU cores, the solution of even a realistic dynamical 1d problem, not to mention the more complicated 2d and 3d problems, could be time consuming using the Fortran programs. Previously, we published auto-parallel Fortran programs \cite{bec2016} suitable for Intel (but not GNU) compiler for solving the GP equation. Hence, a need for the full OpenMP version of the Fortran programs to reduce the execution time cannot be overemphasized. To address this issue, we provide here such OpenMP Fortran programs, optimized for both Intel and GNU Fortran compilers and capable of using all available CPU cores, which can significantly reduce the execution time. 

\noindent\\
{\em Summary of revisions:}
Previous Fortran programs \cite{bec2009} for solving the time-dependent GP equation in 1d, 2d, and 3d with different trap symmetries have been parallelized using the OpenMP interface to reduce the execution time on multi-core processors. There are six different trap symmetries considered, resulting in six programs for imaginary-time propagation and six for real-time propagation, totaling to 12 programs included in BEC-GP-OMP-FOR software package.

All input data (number of atoms, scattering length, harmonic oscillator trap length, trap anisotropy, etc.) are conveniently placed at the beginning of each program, as before \cite{bec2016}. Present programs introduce a new input parameter, which is designated by \texttt{Number\_of\_Threads} and defines the number of CPU cores of the processor to be used in the calculation. If one sets the value 0 for this parameter, all available CPU cores will be used. For the most efficient calculation it is advisable to leave one CPU core unused for the background system's jobs. For example, on a machine with 20 CPU cores such that we used for testing, it is advisable to use up to 19 CPU cores. However, the total number of used CPU cores can be divided into more than one job. For instance, one can run three simulations simultaneously using 10, 4, and 5 CPU cores, respectively, thus totaling to 19 used CPU cores on a 20-core computer.

The Fortran source programs are located in the directory \texttt{src}, and can be compiled by the \texttt{make} command using the \texttt{makefile} in the root directory \texttt{BEC-GP-OMP-FOR} of the software package. The examples of produced output files can be found in the directory \texttt{output}, although some large density files are omitted, to save space. The programs calculate the values of actually used dimensionless nonlinearities from the physical input parameters, where the input parameters correspond to the identical nonlinearity values as in the previously published programs \cite{bec2009}, so that the output files of the old and new programs can be directly compared. The output files are conveniently named such that their contents can be easily identified, following the naming convention introduced in Ref.~\cite{bec2016}. For example, a file named \texttt{<code>-out.txt}, where \texttt{<code>} is a name of the individual program, represents the general output file containing input data, time and space steps, nonlinearity, energy and chemical potential, and was named \texttt{fort.7} in the old Fortran version of programs \cite{bec2009}. A file named \texttt{<code>-den.txt} is the output file with the condensate density, which had the names \texttt{fort.3} and \texttt{fort.4} in the old Fortran version \cite{bec2009} for imaginary- and real-time propagation programs, respectively.
Other possible density outputs, such as the initial density, are commented out in the programs to have a simpler set of output files, but users can uncomment and re-enable them, if needed. In addition, there are output files for reduced (integrated) 1d and 2d densities for different programs. In the real-time programs there is also an output file reporting the dynamics of evolution of root-mean-square sizes after a perturbation is introduced. The supplied real-time programs solve the stationary GP equation, and then calculate the dynamics. As the imaginary-time programs are more accurate than the real-time programs for the solution of a stationary problem, one can first solve the stationary problem using the imaginary-time programs, adapt the real-time programs to read the pre-calculated wave function and then study the dynamics. In that case the parameter \text{NSTP} in the real-time programs should be set to zero and the space mesh and nonlinearity parameters should be identical in both programs. The reader is advised to consult our previous publication where a complete description of the output files is given \cite{bec2016}. A \texttt{readme.txt} file, included in the root directory, explains the procedure to compile and run the programs.

\begin{table}[tp]
\caption{Wall-clock execution times (in seconds) for runs with 1, 6, and 19 CPU cores of different programs using the Intel Fortran (\texttt{ifort}) and GNU Fortran (\texttt{gfortran}) compilers on a workstation with two Intel Xeon E5-2650 v3 CPUs, with a total of 20 CPU cores, and the obtained speedups for 19 CPU cores.}
\label{tab1}
\centering
\begin{tabular}{ccccccccc}
\hline \# of cores
& 1 &1 &6 &6 &19 &19 & 19 & 19 \\
\hline
Fortran & Intel& GNU &Intel &GNU &Intel & GNU & Intel & GNU \\
 & time &time &time & time & time & time & speedup & speedup \\
\hline
imag1d&  52& 60 & 22 & 22& 20 & 22 &  2.6 &  2.7 \\
imagcir&22 & 30 &14 &15 &14 &15 & 1.6& 2.0 \\
imagsph&24& 30 &12 &15 &12 &14 & 2.4 & 2.1 \\
real1d& 205& 345 & 76 &108 & 62 & 86 & 3.3 & 4.0 \\
realcir&  145&  220 & 55& 73 & 48 & 59 & 3.0& 3.7 \\
realsph& 155 &  250& 57 & 76 &  46 &  61& 3.4& 2.7 \\
\hline
imag2d& 255 & 415 &  52 &  84 & 27 & 40 & 9.4& 10.4 \\
imagaxi&  260& 435 & 62 & 105& 30 &  55& 8.7& 7.9  \\
real2d&  325 & 525 & 74 & 107 & 32 &  50& 10.1&  10.5 \\
realaxi& 160& 265 & 35 &  49 & 16 &  24& 10.0& 11.0 \\
\hline
imag3d& 2080 &  2630 &  370 &  550 & 200 & 250 & 10.4& 10.5 \\
real3d&  19500 &  26000 &  3650 &  5600 & 1410 & 2250 &  13.8 & 11.6 \\
\hline
\end{tabular}
\end{table}

\begin{figure}[!t] 
\begin{center}
\includegraphics[width=7.1cm]{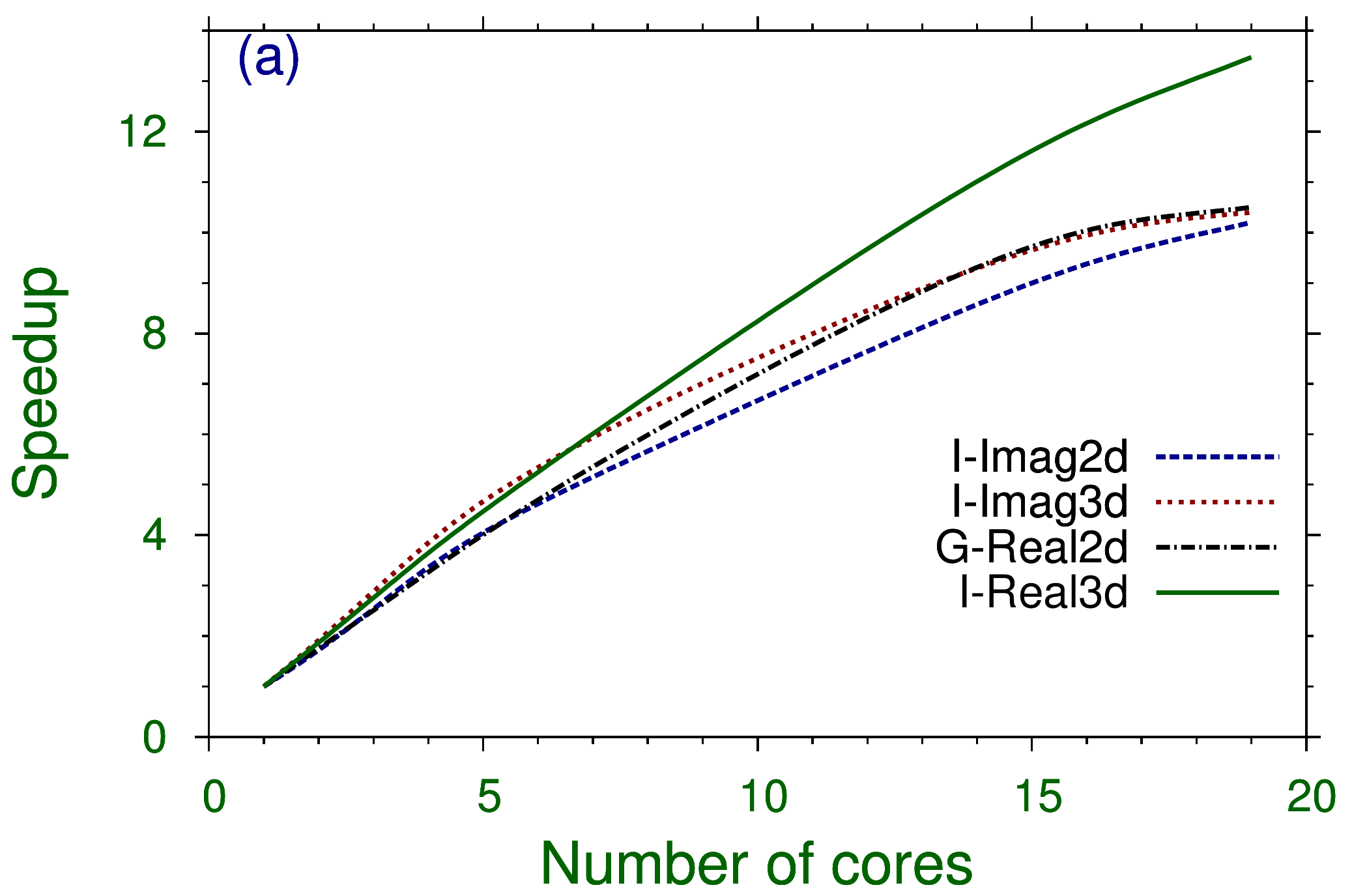}
\includegraphics[width=7.1cm]{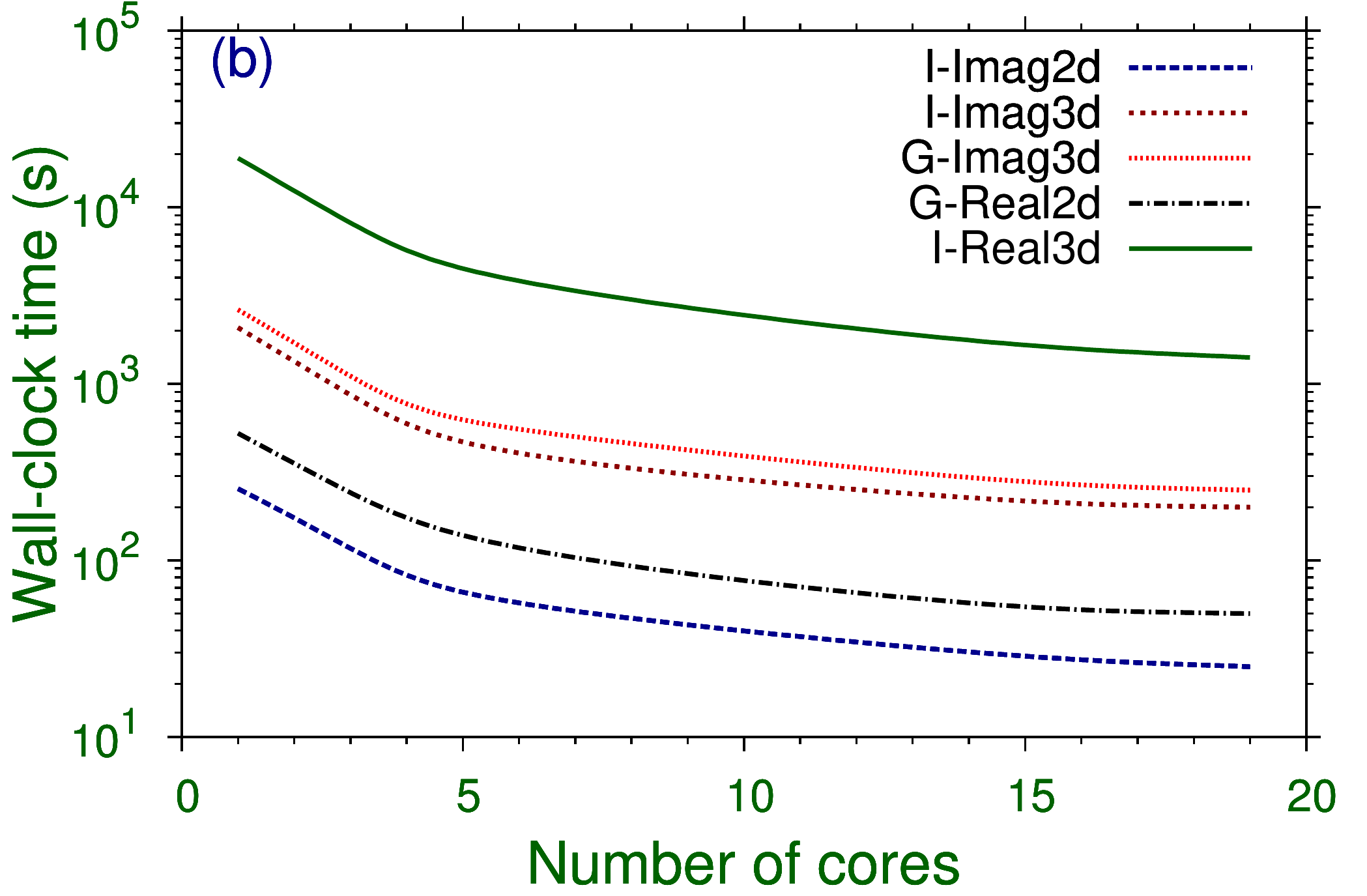}
\caption{(a) Speedup for 2d and 3d programs compiled with the Intel (I) and GNU (G) Fortran compilers as a function of the number of CPU cores, measured on a workstation with two Intel Xeon E5-2650 v3 CPUs. (b) Wall-clock execution time (in seconds) of 2d and 3d programs compiled with the Intel (I) and GNU (G) Fortran compilers as a function of the number of CPU cores.}
\label{fig1}
\end{center}
\end{figure} 

\begin{figure}[!t] 
\begin{center}
\includegraphics[width=8cm]{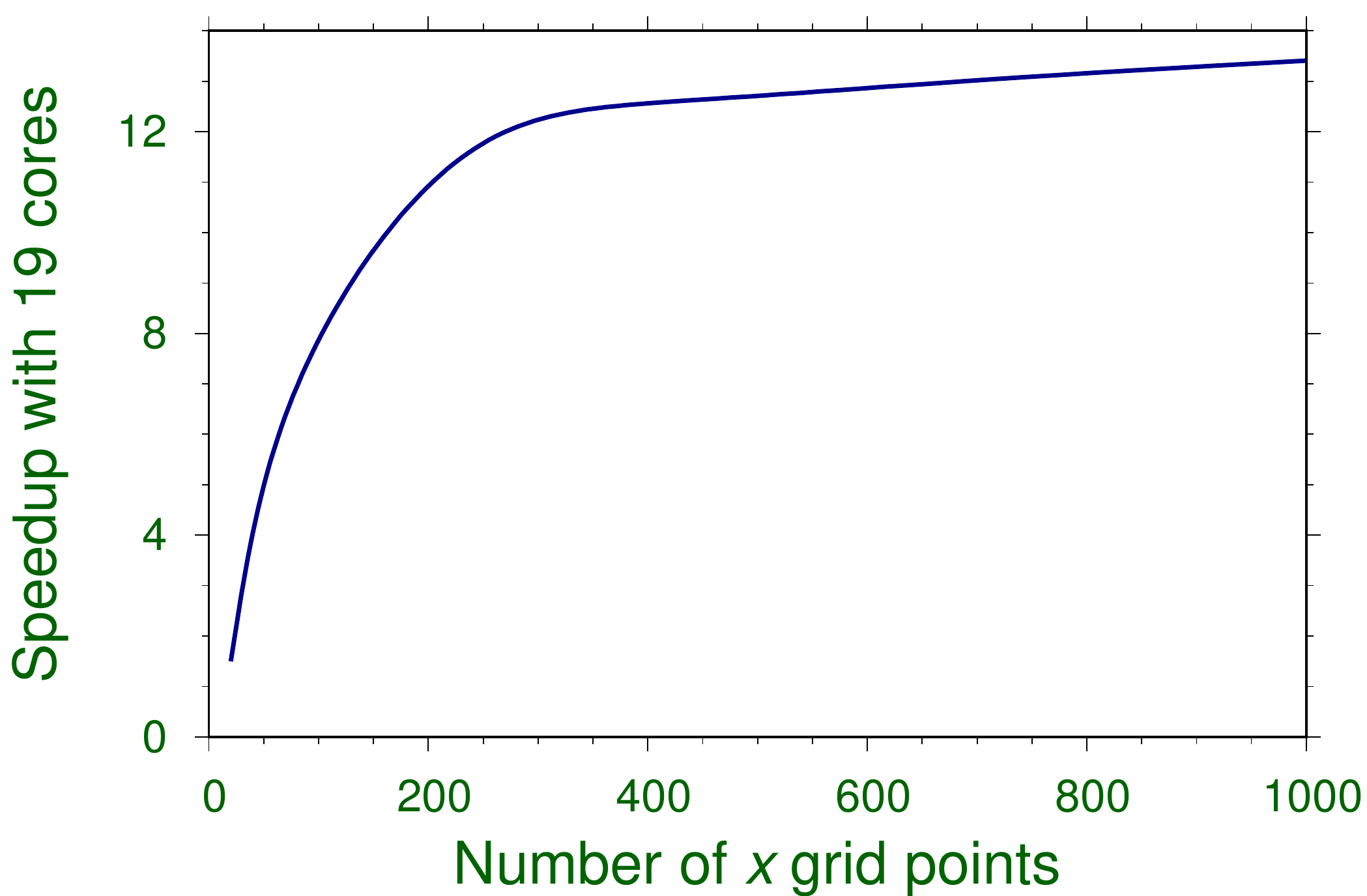} 
\caption{Speedup of real2d-th program, compiled with the Intel Fortran 90 compiler and executed on 19 CPU cores on a workstation with two Intel Xeon E5-2650 v3 CPUs, as a function of the number of spatial discretization points \texttt{NX=NY}. }
\label{fig2}
\end{center}
\end{figure}

We tested our programs on a workstation with two 10-core Intel Xeon E5-2650 v3 CPUs. The parameters used for testing are given in sample input files, provided in the corresponding directory. together with the programs. In Table~\ref{tab1} we present wall-clock execution times for runs on 1, 6, and 19 CPU cores for programs compiled using Intel and GNU Fortran compilers. The corresponding columns ``Intel speedup" and ``GNU speedup" give the ratio of wall-clock execution times of runs on 1 and 19 CPU cores, and denote the actual measured speedup for 19 CPU cores. In all cases and for all numbers of CPU cores, although the GNU Fortran compiler gives excellent results, the Intel Fortran compiler turns out to be slightly faster. Note that during these tests we always ran only a single simulation on a workstation at a time, to avoid any possible interference issues. Therefore, the obtained wall-clock times are more reliable than the ones that could be measured with two or more jobs running simultaneously.  We also studied the speedup of the programs as a function of the number of CPU cores used. The performance of the Intel and GNU Fortran compilers is illustrated in Fig.~\ref{fig1}, where we plot the speedup and actual wall-clock times as functions of the number of CPU cores for 2d and 3d programs. We see that the speedup increases monotonically with the number of CPU cores in all cases and has large values (between 10 and 14 for 3d programs) for the maximal number of cores. This fully justifies the development of OpenMP programs, which enable much faster and more efficient solving of the GP equation. However, a slow saturation in the speedup with the further increase in the number of CPU cores is observed in all cases, as expected.
 
The speedup tends to increase for programs in higher dimensions, as they become more complex and have to process more data. This is why the speedups of the supplied 2d and 3d programs are larger than those of 1d programs. Also, for a single program the speedup increases with the size of the spatial grid, i.e., with the number of spatial discretization points, since this increases the amount of calculations performed by the program. To demonstrate this, we tested the supplied real2d-th program and varied the number of spatial discretization points \texttt{NX=NY} from 20 to 1000. The measured speedup obtained when running this program on 19 CPU cores as a function of the number of discretization points is shown in Fig.~\ref{fig2}. The speedup first increases rapidly with the number of discretization points and eventually saturates.
 
 \noindent\\
{\em Additional comments:} Example inputs provided with the programs take less than 30 minutes to run on a workstation with two Intel Xeon E5-2650 v3 processors (2 QPI links, 10 CPU cores, 25~MB cache, 2.3~GHz).
 
\section*{Acknowledgements}
\noindent
V.L., D.V., and A.B. acknowledge support by the Ministry of Education, Science, and Technological Development of the Republic of Serbia under projects ON171017 and III43007.
P.M. acknowledges support by the Science and Engineering Research Board, Department of Science and Technology, Government of India under project No.~EMR/2014/000644.
S.K.A. acknowledges support by the CNPq of Brazil under project 303280/2014-0, and by the FAPESP of Brazil under project 2012/00451-0. Numerical tests were partially carried out on the PARADOX supercomputing facility at the Scientific Computing Laboratory of the Institute of Physics Belgrade.
\end{small}

\end{document}